\documentclass[singlelespacing]{article}
\usepackage{epsfig}
\usepackage{amssymb}

\begin{document}

\begin{center}

\textbf{{\Large Search for long-lived superheavy eka-tungsten with radiopure ZnWO$_4$ crystal scintillator}}

~~~

P.~Belli$^{a}$,
R.~Bernabei$^{a,b,}\footnote{Corresponding author: rita.bernabei@roma2.infn.it}$,
F.~Cappella$^{c}$,
R.~Cerulli$^{c}$,
F.A.~Danevich$^{d}$,
V.Yu.~Denisov$^{d}$,
A.~d'Angelo$^{e}$,
A.~Incicchitti$^{e}$,
V.V.~Kobychev$^{d}$,
D.V.~Poda$^{d,f}$,
O.G.~Polischuk$^{e,d}$,
V.I.~Tretyak$^{e,d}$

~~~

\it{

$^a$INFN, Sezione di Roma ``Tor Vergata'', I-00133 Rome, Italy

$^b$Dipartimento di Fisica, Universit$\grave{a}$ di Roma ``Tor Vergata'', I-00133 Rome, Italy

$^c$INFN, Laboratori Nazionali del Gran Sasso, I-67100 Assergi (AQ), Italy

$^d$Institute for Nuclear Research, MSP 03680 Kyiv, Ukraine

$^e$INFN, Sezione di Roma ``La Sapienza'', I-00185, Rome, Italy

$^f$Centre de Sciences Nucl\'{e}aires et de Sciences de la Mati\`{e}re, 91405 Orsay, France

}

\end{center}

\begin{abstract}

\noindent
The data collected with a radioactively pure ZnWO$_4$ crystal scintillator (699 g)
in low background measurements during 2130 h at the underground (3600 m w.e.)
Laboratori Nazionali del Gran Sasso (INFN, Italy) were used to set a limit on possible 
concentration of superheavy eka-W (seaborgium Sg, $Z=106$) in the crystal. 
Assuming that one of the daughters in a chain of decays of the initial Sg nucleus decays with 
emission of high energy $\alpha$ particle ($Q_\alpha > 8$ MeV) and analyzing the
high energy part of the measured $\alpha$ spectrum, the limit
$N$(Sg)/$N$(W) $< 5.5\times10^{-14}$ atoms/atom at 90\% C.L. was obtained
(for Sg half-life of $10^9$ yr). 
In addition, a limit on the concentration of eka-Bi
was set by analysing the data collected with a large BGO scintillation bolometer
in an experiment performed by another group [L. Cardani et al., JINST 7 (2012) P10022]:
$N$(eka-Bi)/$N$(Bi) $< 1.1\times10^{-13}$ atoms/atom with 90\% C.L.
Both the limits are comparable with those obtained in recent experiments
which instead look for spontaneous fission of superheavy
elements or use the accelerator mass spectrometry.

\end{abstract}



{\it Keywords:} Superheavy elements; ZnWO$_4$ crystal scintillator; Low background experiment

{\it PACS:} 27.90.+b, 29.40.Mc, 23.60.+e

\section{Introduction}

Possible existence of superheavy elements (SHE) with atomic masses $A \gtrsim 250$ and atomic numbers
$Z \gtrsim 104$ was already discussed in 1950's \cite{Sch57}.
In 1960's, the development of new methods of calculation of the shell model corrections to the liquid drop model
predicted a neutron-rich ``island of stability'' around the double magic $Z=114$ or 126, $N=184$
\cite{Mye66,Sob66,Mel67,Str67,Nil68,Nil69,Fis72}, with
half-life of the nucleus $^{294}_{184}$110 calculated as
$10^8$ yr \cite{Nil69} and $2.5\times10^9$ yr \cite{Fis72}.
Various recent calculations \cite{Smo97,Mol97,Bur98,Kru00,Ber04,Ben01,Sob07,Den07} related to 
different macro-micro and microscopic models predict $N=184$ as the magic number of 
neutrons and $Z=114, 120$ or 126 as the proton magic number for spherical nuclei. 

In experiments on the artificial synthesis of the SHE in fusion of ions with accelerators, more than one hundred
different unstable isotopes with $Z = 104-118$ were created \cite{Hof00,Oga07,Ham13,Tho13}, 
with half-lives from microseconds to hours 
(for $^{268}$Db $T_{1/2}=29^{+9}_{-6}$ h \cite{Oga07}). 
New isotopes with $Z=107-118$ predominantly undergo a chain of $\alpha$ decays followed by 
spontaneous fission (SF) \cite{Ham13}.
Note that the SHE formed in fusion reaction \cite{Hof00,Oga07} are proton-rich 
(or neutron-deficient), and the half-lives of SHE with number of neutrons near 
the magic number 184 are expected to be longer.
 
While maybe only the edge of the ``island of stability'' is reached to-date in the laboratory 
conditions, long-lived SHE probably were produced in explosive stellar events by a sequence of 
rapid neutron captures and $\beta^-$ decays \cite{Sch71} (for current status of our
understanding of SHE nucleosynthesis see e.g. \cite{Pan09,Pet12,Pan13} and refs. therein).
It should be noted that some recent estimations (e.g. \cite{Pet12,Kow10}) are more pessimistic: 
calculations \cite{Pet12}
predict that superheavy nuclei with masses up to $A \simeq 300$ can be produced during star explosions
but they decay through several $\beta$ transformations and spontaneous fission on time scales of days 
not reaching the valley of stability. Such results, however, evidently depend on models (and many 
theoretical parameters) used for description of stars, their evolution and nucleosynthesis during explosion,
as well as on models which predict masses of nuclei and their half-lives respectively to
$\alpha$, $\beta$ decays and spontaneous fission.
Thus, a possibility that long-lived SHE were synthesized in the r-process 
and that they still are present in minor quantities in the Earth materials is not fully excluded.

In 1970's and 1980's, an extensive program to search for SHE in nature was undertaken. 
Hundreds of samples from the ocean floor to the lunar surface were analyzed with sensitivity to mass 
concentration of $10^{-11}-10^{-14}$ g/g (see reviews \cite{Her79,Fle81,Fle83,Kra83,Her14} and refs. therein).
Some hints were obtained on the presence of SHE through their fission in meteorites and in hot spring 
waters from the Cheleken peninsula (Caspian sea), or through their long tracks in olivine crystals
from meteorites (which supposedly belong to SHE with $Z \simeq 110$). 
Giant radioactive halos in minerals were also considered as possibly caused
by radiation damage due to $\alpha$ particles with energy $10-15$ MeV emitted 
from the central inclusion by SHE with $Z \simeq 120$ \cite{Gen70,Gen76}.
While the old results are summarized in \cite{Her79,Fle81,Fle83,Kra83,Her14}, we will give below
details on more recent investigations.

In the measurements of Marinov et al. with natural gold using inductively coupled plasma sector field 
mass spectrometry, superheavy isotopes with atomic masses $A=261$ and 265 and concentration
$\delta = (1-10)\times10^{-10}$ atoms/atom relatively to Au were found \cite{Mar09}. It was proposed that they are 
most probably isotopes of roentgenium Rg (eka-Au, $Z=111$).
In similar studies on natural thorium, superheavy nuclei with $A=292$ and $\delta \simeq 10^{-12}$ atoms/atom
relatively to $^{232}$Th were found, interpreted as possible existence of long-lived eka-Th 
($Z=122$) \cite{Mar10}.
It should be noted, however, that the evidence on the existence of SHE with $A=261, 265$
in natural Au was not confirmed in sensitive searches with accelerator mass spectrometry (AMS)
of the Vienna group 
\cite{Del11a} where few orders of magnitude lower limits of $\delta < 3\times10^{-16}$ atoms/atom were obtained. 
Similarly, only limit of $\delta < 4\times10^{-15}$ atoms/atom was found for abundance of SHE with $A=292$ 
in Th \cite{Del10} (see also comments in \cite{Mar11}).
The AMS searches of the same group for SHE in natural Pt, Pb and Bi also gave only the limits \cite{Del11b}:
i) $\delta < (0.9-2)\times10^{-15}$ atoms/atom for eka-Pt (Ds, $Z=110$) in Pt (for SHE atomic masses $A = 288-295$);
ii) $\delta < (2-6)\times10^{-14}$ atoms/atom for eka-Pb (Fl, $Z=114$) in Pb (for $A = 292,293,295-297,299$);
iii) $\delta < (5-30)\times10^{-13}$ atoms/atom for eka-Bi ($Z=115$) in Bi (for $A = 293-300$).

Recently the AMS technique was also used in the searches of the Garching group 
for SHE with $292 \le A \le 310$ in samples of Os,
Pt and PbF$_2$, where only limits were established in the range of 
$1.5\times10^{-16} - 4.1\times10^{-14}$ atoms/atom \cite{Lud12}.

In the OLIMPIYA experiment \cite{Bag13}, in the analysis of $\simeq6000$ tracks from cosmic rays accumulated 
during $(1-3)\times10^8$ yr in $\simeq170$ olivine crystals extracted from meteorites, three tracks were
found which could belong to nuclei with $105 < Z < 130$. A more accurate estimation for one of them
gave $Z=119^{+10}_{-6}$ at 95\% C.L. \cite{Bag13,Ale13}.

Analyzing recent theoretical calculations, Oganessian concluded \cite{Oga07,Oga96} that the most
stable superheavy nuclide could be not eka-Pb with $Z=114$ but nuclides with $A \simeq 290$
and $Z=106,107,108$ 
(eka-W, eka-Re, eka-Os, respectively). It was expected that they will
decay through chain of $\beta$ and $\alpha$ decays which may end with spontaneous fission.
It was proposed \cite{Oga96} to use thin foils ($\sim 0.1$ mg/cm$^2$) to register expected 
$\alpha$ particles (this immediately constrains masses of samples on the scale of grams).

A search for eka-Os (Hs, $Z=108$) was carried out in the experiment SHIN (SuperHeavy In Nature) deep
underground (4800 m w.e., to suppress background from neutrons created by cosmic muons) 
in the Laboratoire Souterrain de Modane (France) \cite{Svi09}. 
It was expected that either the initial Hs nucleus or its daughter created in chain of $\alpha$
decays will finally decay through SF.
The average number of neutrons per fission is equal to $\bar{\nu} \simeq 4.5$ for $Z=104$ and 
$\bar{\nu} \simeq 6$ for $Z=108$, very different from that from possible background from SF of 
$^{238}$U ($\bar{\nu} \simeq 2$). An Os sample with mass of 550 g was measured during 3 yr with
a neutron detector which consisted of 60 $^3$He counters placed in 4 rings around the sample.
Few events with multiplicity $\nu \geq 3$ were observed, but only a limit on mass concentration 
of eka-Os in Os $\delta \leq 10^{-14}$ g/g was set (with the standard assumption that the 
half-life of eka-Os is
$T_{1/2}=10^9$ yr). Afterwards the Os sample was changed with a Xe sample
(140 g) to search for superheavy homolog of Xe; 
the sensitivity was estimated as $\delta \simeq 10^{-13}$ g/g.

In the present work, we use an alternative approach: instead to search for the spontaneous fission,
we look for high energy $\alpha$ particles ($Q_\alpha > 8$ MeV) possibly emitted in a chain of 
decays of eka-W (seaborgium Sg, $Z=106$) and registered by a radioactively pure
ZnWO$_4$ crystal scintillator working in a low background installation at the
Laboratori Nazionali del Gran Sasso (LNGS) of the INFN (Italy) at a depth of 3600 m w.e.
Chemical properties of seaborgium are similar to those of tungsten \cite{Sch97,Per13,Eve14}, and it is expected
that Sg in some amount follows W in processes of chemical purification and growth of a ZnWO$_4$ 
crystal \footnote{Because of the actinoid contraction, the atomic and ionic radii of trans-actinoids in 
the 7-th period are expected to be close to the radii of their homologs in the previous period
of the periodic table, in the same way as the lanthanoid contraction makes the radii of
pre- and post-lanthanoid homologs to be similar. Thus, one should expect that the
trans-actinoids would easily substitute their homologs in crystal lattices.}.
Such a technique, when a source of radiation is embedded in a detector (``source = detector'' approach),
allows to reach practically the 100\% efficiency in the registration of the process
and to use samples with masses on the scale of kg.
Measurements with several ZnWO$_4$ detectors were
carried out earlier, devoted mainly to the search for double beta decay processes in Zn and W isotopes 
\cite{Bel09,Bel11a}; information on their radiopurity was also published \cite{Bel11b}.
Here we reanalyze the data for possible presence of SHE in ZnWO$_4$ crystals.

\section{Experiment and data processing}

The detailed description of the set-up with ZnWO$_4$ crystal scintillators and its
performances have been discussed in \cite{Bel09,Bel11a,Bel11b}.
Four ZnWO$_4$ detectors were measured (with mass of 117, 141, 239 and 699 g). 
While their characteristics were very similar, the largest of them was also found to be the most
radioactively pure \cite{Bel11b}, and in the following we will consider only it in more detail. 
Here we recall the main features of the measurements.

The ZnWO$_4$ crystal scintillator ($\oslash44\times55$ mm, mass of 699 g)
was grown by the Czochralski method.
It was fixed inside a cavity of $\oslash47\times59$ mm in the central part of a polystyrene
light-guide 66 mm in diameter and 312 mm in length. The cavity was
filled up with high purity silicone oil. The light-guide was
optically connected on opposite sides by optical couplant to two
low radioactivity EMI9265--B53/FL 3'' photomultipliers (PMT). The
light-guide was wrapped by PTFE tape.

The detector has been installed deep underground ($\simeq3600$ m
w.e.) in the low background DAMA/R\&D set-up at the LNGS. 
It was surrounded by Cu bricks and sealed in a low
radioactive air-tight Cu box continuously flushed with high purity
nitrogen gas to avoid
presence of residual environmental Radon. The copper box was
surrounded by a passive shield made of 10 cm of high purity Cu, 15
cm of low radioactive lead, 1.5 mm of cadmium and 4/10 cm
polyethylene/paraffin to reduce the external background. The whole
shield has been closed inside a Plexiglas box, also continuously
flushed by high purity nitrogen gas.

An event-by-event data acquisition system accumulates the
amplitude and the arrival time of the events. The time profile of the sum of the
signals from the PMTs was also recorded with a sampling frequency of
20 MS/s over a time window of 100 $\mu$s by a 8 bit transient
digitizer (DC270 Acqiris).
A CAMAC system was used to manage 
the triggers and the ADCs, introducing a rather long dead time of about 
26 ms (in particular, because of storing the time profiles on a hard disk).

The energy scale and energy resolution of the ZnWO$_4$ detector have been
measured with $^{22}$Na, $^{133}$Ba, $^{137}$Cs, $^{228}$Th and
$^{241}$Am $\gamma$ sources. 
The energy resolution of the ZnWO$_4$ detector for $\gamma$ quanta
is well described as FWHM$_\gamma$(keV) = $\sqrt{13.96(47) \times E_\gamma}$, 
where $E_\gamma$ is the energy of $\gamma$ quanta in keV \footnote{Because
$\gamma$ rays interact with matter by means of the energy transfer to 
electrons, the same is applicable also for $\beta$ particles;
we use abbreviation``$\gamma(\beta)$'' in such cases in the following.}.

It is known that light yield for $\alpha$ particles in scintillators
is lower than that for $\gamma$ quanta ($\beta$ particles) of the same energy
\cite{Bir64}. 
The ratio of the $\alpha$ peak position in the energy scale measured
with $\gamma$ sources to the real energy of the $\alpha$ particles 
for $E_{\alpha}>2$ MeV is described as
$\alpha/\beta=0.074(16)+0.0164(40)\times E_{\alpha}$, where
$E_{\alpha}$ is the energy of $\alpha$ particles in MeV.
The energy resolution for $\alpha$ particles is:
FWHM$_\alpha$(keV) = $33+0.247 \times E_\alpha^\gamma$,
where $E_\alpha^\gamma$ is energy of $\alpha$ particles in $\gamma$ 
scale (in keV) \footnote{The energy resolution for $\alpha$ particles is worse than
that for $\gamma$ quanta due to the dependence of the $\alpha/\beta$ ratio on the
direction of the $\alpha$'s relatively to the ZnWO$_4$ crystal axes \cite{Dan05}.}.

The scintillation signals induced in ZnWO$_4$ by $\alpha$ particles
are different from those caused by $\gamma(\beta)$ particles (signals from $\alpha$'s
are shorter) \cite{Dan05}. This difference allows one to discriminate
$\gamma(\beta)$ events from $\alpha$ events. 
We used for this purpose the optimal filter method \cite{Gat62}. 
For each signal,
the numerical characteristic of its shape (shape indicator, $SI$)
was defined as $SI=\sum f(t_k)\times P(t_k)/\sum f(t_k)$, where
the sum is over the time channels $k$, starting from the origin of
signal and up to 50 $\mu$s; $f(t_k)$ is the digitized amplitude
(at the time $t_k$) of a given signal. The weight function $P(t)$
was defined as: 
$P(t)=[f_\alpha (t)-f_\gamma (t)]/[f_\alpha(t)+f_\gamma (t)]$, 
where $f_\alpha (t)$ and $f_\gamma (t)$ are
the reference pulse shapes for $\alpha$ particles and $\gamma$
quanta, respectively, obtained by summing up the shapes of a few
thousand $\gamma$ or $\alpha$ events. 
The scatter plot of the
shape indicator versus energy for the data of the low background
measurements with the ZnWO$_4$ detector during 2130 h is shown in Fig.~1. 
The distributions of the $SI$ values for the $\alpha$ and $\gamma(\beta)$ events 
are well described by Gaussian functions, with center and width dependent on 
energy (the $SI$ distributions for events in the energy interval $800-900$ keV
are shown in Inset of Fig.~1). 
Using these dependencies, $\pm2\sigma$ contours were calculated where 95\% of 
the corresponding events are contained. 
Detailed analysis of the obtained $\gamma(\beta)$ and $\alpha$ spectra is
given in \cite{Bel09}; clusters within the $\gamma(\beta)$ limits belong to
$^{40}$K, $^{65}$Zn, $^{208}$Tl, $^{214}$Bi, and alpha events are related 
mainly with $^{238}$U and its daughters.
More details on measurements and data processing can
be found in \cite{Bel09,Bel11a,Bel11b}.

One-dimensional energy spectrum of $\alpha$
events inside the $\pm2\sigma$ contour is shown in Fig.~2 where the real energies of the
$\alpha$ particles were obtained using the $\alpha/\beta$ ratio given above.

\nopagebreak
\begin{figure}[htb]
\begin{center}
\mbox{\epsfig{figure=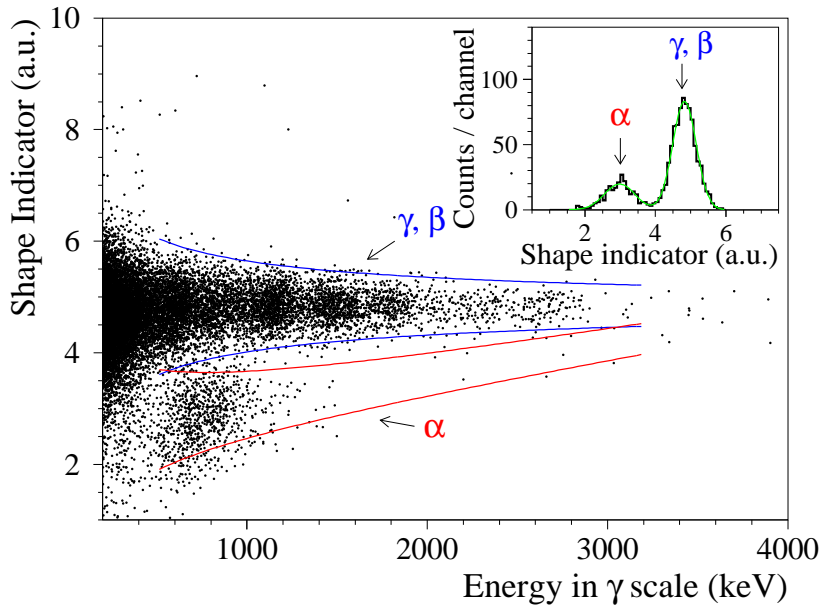,height=8.0cm}} 
\caption{Scatter plot of the shape indicator (see text) versus energy 
for 2130 h background measurements with the ZnWO$_4$ crystal scintillator
having a mass of 699 g. The contours give regions where 95\% of $\alpha$ or 
$\gamma(\beta)$ events are concentrated.
(Inset) The $SI$ distributions of the events in the energy interval 800 -- 900 
keV. The $\alpha$ and $\gamma(\beta)$ distributions are fitted by Gaussian 
functions (solid line).}
\end{center}
\end{figure}

\nopagebreak
\begin{figure}[htb]
\begin{center}
\mbox{\epsfig{figure=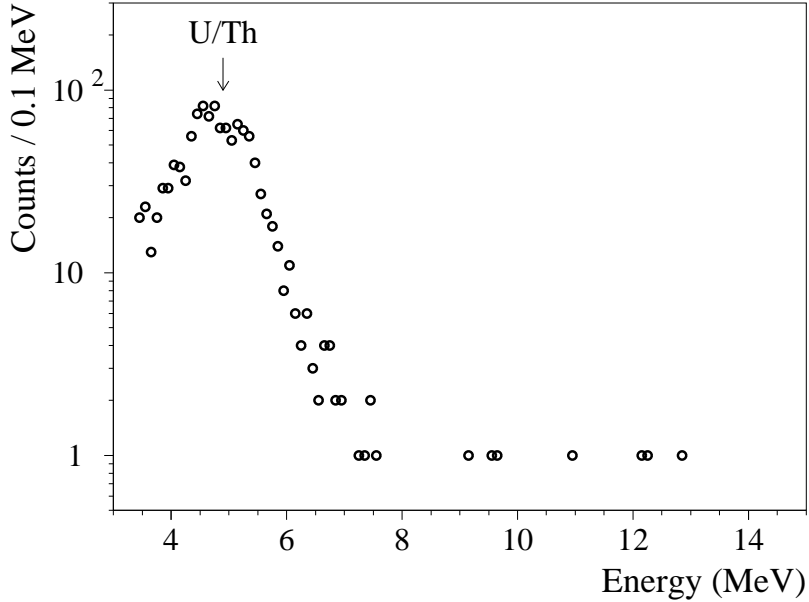,height=8.0cm}} 
\caption{One-dimensional energy spectrum of $\alpha$ particles registered 
by the 699 g ZnWO$_4$ detector during 2130 h.}
\end{center}
\end{figure}

\section{Limit on long-lived Sg in ZnWO$_4$ crystal}

As it was shown in \cite{Sch97,Per13,Eve14}, the chemical properties of superheavy Sg are
similar to those of W; thus, one could expect that long-lived Sg follows W in the processes
of chemical separation and growth of the ZnWO$_4$ crystals, and could also be present at some
amount in the ZnWO$_4$ detector. 
If in a chain of decays of initial Sg nucleus an $\alpha$ decay with high 
$Q_\alpha$ value occurs, we can see it in the high energy part of the spectrum presented in Fig.~2.

The superheavy isotopes can decay through emission of $\beta^-$ particle or by
electron capture EC (or $\beta^+$ decay), $\alpha$ decay or spontaneous fission; cluster decay also
starts to be important at higher $Z$ values (see recent reviews \cite{Sob07,Poe13}).
In general, energy release $Q_\alpha$ in $\alpha$ decay is increasing with the increase
of the atomic number $Z$; in accordance with \cite{Sil12}, experimental $Q_\alpha$ values
for superheavy elements with $Z=102-107$ lay in the interval 7.8 -- 10.6 MeV.

The decay of superheavy Sg ($Z=106$) will lead to independent events in the high energy part of
the $\alpha$ spectrum in the following scenario:

(1) Long-lived initial Sg nucleus decays through $\beta^-$ channel (thus creating a nucleus
with $Z=107$) or through EC/$\beta^+$ (leading to a nucleus with $Z=105$) or through 
$\alpha$ decay (resulting in $Z=104$). The energy of this first $\alpha$ decay should be 
quite modest ($\simeq 4-6$ MeV, similar to that in decay of $^{232}$Th and $^{235,238}$U) to be consistent
with the large $T_{1/2}$. Such a low energy, however, does not allow to distinguish it from
the decays in usual U/Th chains present in trace amount in any material;

(2) The created nucleus (or one of its daughters) decays with emission of high energy $\alpha$ 
particle ($Q_\alpha > 8$ MeV). The energy threshold of 8 MeV allows to cut off
contributions from $\alpha$ decays in U/Th chains because all nuclides in these chains
have $Q_\alpha < 8$ MeV, with the exception of $^{212}$Po with $Q_\alpha = 8.954$ MeV \cite{ToI98};

(3) This daughter nuclide with $Q_\alpha > 8$ MeV, nevertheless, should live long enough
to be registered outside of the dead time of the data acquisition system after the preceding 
decay (in our measurements it is 26 ms \footnote{It should be noted that 
very fast chains of decays also can be found: by analysis of
the time profile of events, which is recorded in our measurements during 100 $\mu$s.
However, this is outside of the present work.}).
This is also realistic because many of the experimentally discovered to-date 
SHE isotopes with $Q_\alpha > 8$ MeV have $T_{1/2}$ in the range of seconds (or larger) \cite{Ham13,Sil12}.
The theoretical predictions \cite{Cho08,Den10,War12} also confirm this assumption. 

It is difficult to recognize exactly initial Sg nuclide and its chain of decays which fulfill all the above
conditions. One of the pretendents could be $^{290}_{106}$Sg with half-lives relatively to
spontaneous fission and $\alpha$ decay estimated in \cite{Smo97} as: 
$T_{1/2}^{\textrm{sf}} = 1.3 \times 10^6$ yr and 
$T_{1/2}^\alpha = 1.6 \times 10^{11}$ yr. Only limit for $\beta$ decay was given in \cite{Mol97}:
$T_{1/2}^\beta > 100$ s, and this nucleus is quite stable in respect to cluster radioactivity:
$T_{1/2}^{\textrm{c}} \simeq 10^{20}$ yr \cite{Poe11}.
However, the expected $T_{1/2}$ values for all possible decay channels of all potential daughters 
of $^{290}_{106}$Sg (and its neighbours) are not available in the literature. One has also to
remember that all theoretical estimations strongly depend on the models (and their parameters) 
used for calculation of nuclear masses and half-lives for $\alpha$, $\beta$/EC, SF and cluster
decay channels (just for example, in the above mentioned works $T_{1/2}^\alpha$ of
$^{290}_{106}$Sg was estimated as:
$1.6 \times 10^{11}$ yr \cite{Smo97}, 
$9.1 \times 10^{8}$ yr \cite{Mol97}), 
$1.2 \times 10^{8}$ yr \cite{Cho08}.
Taking into account such theoretical uncertainties, experimental investigations should be 
performed too.

If conditions (1) -- (3) are fulfilled, we can use the data collected in the low background measurements with
the ZnWO$_4$ scintillator during $t=2130$ h to restrict the presence of long-lived
Sg in the crystal. As one can see in Fig.~2, we have $S=7$ events in the energy interval 8 -- 15
MeV. Some of them (at lower energies) could be related with $^{212}$Po, some of them
(at higher energies) could be $\gamma(\beta)$ events (see Fig.~1)
or the so-called $^{212}$Bi-$^{212}$Po events \cite{Bel11a}, 
but here we conservatively take all of them to derive limit on number $N$ of Sg nuclei. 
For time of measurements $t << T_{1/2}$ one can use relation: 
$S = \ln 2 \cdot \varepsilon \cdot N \cdot t / T_{1/2}$, where  
$\varepsilon$ is the efficiency to register $\alpha$ decay in the ZnWO$_4$ detector.
It is clear that the probability to absorb $\alpha$ particle in the large (699 g) crystal
is practically equal to 1. We cannot estimate loss of efficiency due to the dead time
of the data acquisition because half-life of $\alpha$ decaying nucleus is unknown.
Thus, in the following we will use only $\varepsilon = 0.95$ determined by the cuts in the
selection of the $\alpha$ events (see Fig.~1).
For the half-life of the initial Sg nucleus, we will adopt the value: $T_{1/2}=10^9$ yr
which is a standard assumption in the SHE searches in nature \cite{Fle83,Svi09}.
With the measured number of $S = 7$ events and very conservatively supposing 0 background, 
the number of decays is limited by $\lim S < 11.77$ at 90\% C.L. in accordance with \cite{Hel83}.
With all the values given above,
one obtains the limit on the number of Sg nuclei as:
$\lim N$(Sg) $< 7.4\times 10^{10}$. Since the 
number of W nuclei in the 699 g ZnWO$_4$ crystal is $N$(W) $= 1.34\times10^{24}$,
one gets the concentration of Sg in ZnWO$_4$ relatively to W:
\begin{equation}
N \textrm{(Sg)} /N \textrm{(W)} < 5.5\times10^{-14}~\textrm{atoms/atom at 90\% C.L.}
\end{equation}
This value is comparable with the sensitivity reached in the searches for eka-Os in the SHIN 
experiment ($\delta \leq 10^{-14}$ g/g) \cite{Svi09}.

However, the above limit has a drawback: it is given relatively not to natural W but relatively to
W in our ZnWO$_4$, i.e. to W which passed processes of extraction from initial W-containing
ores (commercially important minerals are mainly scheelite CaWO$_4$ and wolframite (Fe,Mn)WO$_4$), 
purification and growth of the ZnWO$_4$ crystal. 
It is difficult to estimate possible losses of superheavy Sg in all these processes but one can 
try to do this indirectly, using information on the lighter homolog of W: molybdenum. 
The following typical values can be found for content of Mo in W supplied by different companies:
150 ppm for 99.95\% W (GoodFellow \cite{GoodFellow}),
30 ppm for 99.97\% WO$_3$ and up to 3000 ppm for W (Wolfram \cite{Wolfram}),
100 ppm for 99.998\% (except Mo) WO$_3$ (Alfa Aesar \cite{Alfa}),
12 ppm for 99.97\% W (PLANSEE \cite{PLANSEE}).
If we conservatively assume 10 ppm content of Mo in WO$_3$ powder used in production of 
the ZnWO$_4$ crystal and 1\% content of Mo in an initial W-containing mineral
(both these values are unknown for our ZnWO$_4$),
we can estimate reduction factor of Mo during extraction and purification processes as $\sim 10^3$.

As regards additional losses of Mo during the growth of the ZnWO$_4$ crystal, they can be estimated as 
essentially lower. For this, we can use available information on ZnMoO$_4$ crystal (which is very close 
to ZnWO$_4$) \cite{Ber14}: content of W in initial MoO$_3$ was 200 ppm, and in the grown ZnMoO$_4$
it was 190 ppm, so no W was lost
due to the very low segregation of W in the system of melt and solid ZnWO$_4$ crystal. 

Using these estimations, the limit (1) recalculated for natural W can be up to 3 orders of magnitude 
worse. To avoid this drawback, in future we plan to search for superheavy Sg 
in similar measurements with natural scheelite (CaWO$_4$) crystal.

\section{Potentiality of scintillators to search for other SHE}

The approach to look for high energy $\alpha$'s from the decay of natural SHE (or its
daughters), embedded in a detector, can be also used to search for other SHE.

As an example, Bi$_4$Ge$_3$O$_{12}$ (BGO) scintillators or scintillating bolometers can be used in searches
for superheavy eka-Bi ($Z=115$). In particular, in the data of ref. \cite{Car12}
collected during 455 h with a 891 g BGO scintillating bolometer in a low background set-up
underground at LNGS, one can see (Figs.~2 and 3 of \cite{Car12}) 3 $\alpha$ events in the
energy interval $9.5-10$ MeV (while one could expect no events after the energy of
the most energetic in the U/Th chains $\alpha$ particles from $^{212}$Po with $Q_\alpha = 8.954$ MeV).
The nature of these events should be
investigated additionally (they could be e.g. pile-ups of two $\alpha$ signals or BiPo events) but here
we just give a conservative limit on the presence of eka-Bi in the BGO. In accordance with
the procedure given above one gets 
the limit on the number of events: $S < 6.68$ at 90\% C.L. \cite{Hel83}.
Supposing once more $T_{1/2}=10^9$ yr for eka-Bi, this number of decays during 455 h
corresponds to a number of eka-Bi nuclei as $N$(eka-Bi) = $1.9\times10^{11}$.
The number of the Bi nuclei in the 891 g BGO is: $N$(Bi) = $1.7\times10^{24}$, and thus the
limit on the eka-Bi concentration in BGO is $\delta < 1.1\times10^{-13}$ atoms/atom relatively to Bi.
This is better than the limits obtained in recent searches with the accelerator
mass spectrometry $\delta < (5-30)\times10^{-13}$ atoms/atom (for $A = 293-300$) \cite{Del11b},
demonstrating the good potentiality of the approach considered here to search for SHE.
It should be remembered, however, that this limit is also obtained relatively not to natural Bi but for
Bi in the BGO crystal after processes of Bi extraction, purification and the BGO crystal growth.
While one could expect a reduction factor for natural Bi similar to that in case of W, this
question has to be studied additionally.

\section{Conclusions}

The data collected with a radioactively pure ZnWO$_4$ crystal scintillator in low background
measurements during 2130 h at LNGS were used to set a limit on possible concentration of
superheavy eka-W (seaborgium Sg, $Z=106$) in the crystal. 
Assuming that one of the daughters in a chain of decays of the initial Sg nucleus decays with 
emission of high energy $\alpha$ particles ($Q_\alpha > 8$ MeV) and that half-life
of the long-lived Sg is $10^9$ yr (that is standard assumption in this field 
\cite{Fle83,Svi09}), we obtained the limit on Sg concentration as:
$N$(Sg)/$N$(W) $< 5.5\times10^{-14}$ atoms/atom at 90\% C.L.
This is comparable with the limit $\simeq 10^{-14}$ atoms/atom obtained for concentration
of eka-Os in Os in the recent SHIN experiment \cite{Svi09}. 
One should note that the detection of spontaneous fission, as in SHIN, and the detection of
high energy $\alpha$ particles, as here, are complementary approaches in the searches for
SHE in nature.

The limit on the concentration of eka-Bi in Bi: $N$(eka-Bi)/$N$(Bi) $< 1.1\times10^{-13}$ atoms/atom
was also set from the measurements with a large BGO scintillation bolometer of \cite{Car12}. 
This is comparable with limits obtained with the accelerator
mass spectrometry $\delta < (5-30)\times10^{-13}$ atoms/atom (for $A = 293-300$) \cite{Del11b}.

Both the limits for eka-W and eka-Bi were obtained relatively not to natural W or Bi but for
W and Bi present in ZnWO$_4$ and BGO crystals, i.e. after processes of extraction, purification and
crystals growth. Recalculated for natural W and Bi, limits can be lower by a factor of $\sim 10^3$.

The approach followed in this work can be also used in searches for other superheavy elements
as eka-Tl in NaI(Tl), eka-Pb in PbWO$_4$, and superheavy homolog of Xe in xenon.
It should be noted that detection of $\alpha$ particles with $Q_\alpha>8$ MeV cannot definitively 
point to a specific isotope (e.g. eka-W or eka-Bi). However, if events with shape of 
scintillation signal typical for $\alpha$'s and high energy will be
persistently registered in convincing amounts and alternative explanations 
(like Bi-Po events, pile-ups) will be absent, this would be an indication
on presence of SHE and necessity to involve also other methods for
investigation. 

We are grateful to M. Velazquez and Ya. Vasiliev for useful discussions on impurities in 
ZnMoO$_4$, ZnWO$_4$ and BGO crystals. 
We would like to thank anonymous referees for their remarks which allowed to improve 
our paper.

\end{document}